\documentclass[a4paper,12pt,english]{article}

\usepackage{amsfonts,bm,amssymb,array,babel,cite}
\usepackage[mathscr]{euscript}
\usepackage[all,color]{xy}
\usepackage{relsize}
\usepackage{tikz}
\usepackage{url}
\usepackage{amsmath,amsthm,bm}
\usepackage{accents}
\usepackage[]{graphicx}
\usepackage[T1]{fontenc}

\usepackage{empheq}
\usepackage{color}
\usepackage{mathtools}

\makeatletter

\newcommand{\dd}{\mathrm{d}}

\newcommand{\bbm}{\left(\begin{matrix}}
\newcommand{\ebm}{\end{matrix}\right)}
\newcommand{\beq}{\begin{eqnarray}}
\newcommand{\eeq}{\end{eqnarray}}

\makeatother

\newcommand{\sfrac}[2]{{\textstyle\frac{#1}{#2}}}

\newcommand{\be}{\begin{equation}}
\newcommand{\ee}{\end{equation}}

\newcommand{\beqa}{\begin{eqnarray}}
\newcommand{\eeqa}{\end{eqnarray}} 
\def\nn{\nonumber} \def \bea{\begin{eqnarray}} \def\eea{\end{eqnarray}}

\newcommand{\barr}{\begin{array}}
\newcommand{\earr}{\end{array}}
\numberwithin{equation}{section}

\def\a{\alpha}  
  
 \def\d{\delta}

\def\l{\lambda}   \def\m{\mu}
\def\n{\nu}    
\def\s{\sigma}  \def\t{\tau} 
   
\def\u{v}

 \def\one{\mbox{1 \kern-.59em {\rm l}}}

\def\bi{\begin{itemize}} \def\ei{\end{itemize}}

\def\n2{\bm\nabla^2}

\def\mass{\footnotesize{\textrm M}}
\begin{document}

\makeatother

\parindent=0cm

\renewcommand{\title}[1]{\vspace{10mm}\noindent{\Large{\bf

#1}}\vspace{8mm}} \newcommand{\authors}[1]{\noindent{\large

#1}\vspace{5mm}} \newcommand{\address}[1]{{\itshape #1\vspace{2mm}}}

\begin{titlepage}


\begin{center}

\vskip 3mm

\title{ {\Large Newton-Cartan Gravity and Torsion}
 }

 \authors{Eric Bergshoeff$^{\a,}$$\footnote{e.a.bergshoeff@rug.nl}$,
 Athanasios {Chatzistavrakidis}$^{\a,\beta,}$$\footnote{a.chatzistavrakidis@gmail.com}$, \\
 Luca Romano$\footnote{ lucaromano2607@gmail.com}$
 and Jan Rosseel$^{\gamma,}$$\footnote{rosseelj@gmail.com}$

 }

\vskip 1mm

\address{
$^{\a}$Van Swinderen Institute for Particle Physics and
Gravity, University of Groningen, \\
Nijenborgh 4, 9747 AG Groningen, The Netherlands

\bigskip

$^\beta$Division of Theoretical Physics,
 Rudjer Bo$\check s$kovi\'c Institute, \\
 Bijeni$\check c$ka 54, 10000  Zagreb, Croatia

\bigskip

$^\gamma$Faculty of Physics, University of Vienna,\\
Boltzmanngasse 5, A-1090, Vienna, Austria

}

\smallskip

\end{center}

\vskip 1mm

 \begin{center}
\textbf{Abstract}
\vskip 3mm
\begin{minipage}{14cm}%
We compare the gauging of the Bargmann algebra, for the case of arbitrary torsion, with the result that one obtains from a null-reduction of General Relativity. Whereas the two procedures lead to the same result for Newton-Cartan geometry with arbitrary torsion, the null-reduction of the Einstein equations necessarily leads to Newton-Cartan gravity with zero torsion. We show, for three space-time dimensions,  how Newton-Cartan gravity with arbitrary torsion can be obtained by starting from a Schr\"odinger field theory with dynamical exponent $z=2$ for a complex compensating scalar and next coupling this field theory to a $z=2$ Schr\"odinger geometry with arbitrary torsion. The latter theory can be obtained from either a gauging of the Schr\"odinger algebra, for arbitrary torsion, or from a null-reduction of conformal gravity.

\end{minipage}

 \end{center}

\end{titlepage}

\tableofcontents


\section{Introduction}
\label{intro}

Usually, when discussing Newton-Cartan (NC) geometry and gravity, Newtonian causality is incorporated by imposing that the space-time manifold admits a one-form $\tau_\mu$, called the time-like Vierbein\,\footnote{Most of this paper applies to any space-time dimension. We will therefore from now on use the word Vielbein instead of Vierbein.}, whose curl is constrained to vanish. The vanishing of the curl of $\tau_\mu$ is often referred to as the `zero torsion condition' and implies the existence of an absolute time in the space-time geometry. Indeed, using the one-form $\tau_\mu$ one can define the time difference $T$ between two events as
\begin{equation}
  T = \int_{\mathcal{C}} \dd x^\mu \, \tau_\mu \,,
\end{equation}
where $\mathcal{C}$ is a path connecting the two events. The zero torsion condition implies that the time difference $T$ is independent of the path $\mathcal{C}$ connecting the two events and can thus indeed be identified with an absolute time. Alternatively, the zero torsion condition allows one to express $\tau_\mu$ as the derivative of a single scalar field $\tau(x)$:
\begin{equation}\label{condition1}
\partial_\mu\tau_\nu-\partial_\nu\tau_\mu=0\hskip .5truecm \Rightarrow\hskip .5truecm \tau_\mu=\partial_\mu\tau\,.
\end{equation}
Since then
\begin{equation}
  T = \int_{\mathcal{C}} \dd x^\mu \tau_\mu = \int_{\mathcal{C}} \dd \tau \,,
\end{equation}
one sees that the absolute time $t$ can be identified with this function $\tau(x)$:
\begin{equation}
 \tau(x)=t\hskip .5truecm\Rightarrow\hskip .5truecm \tau_\mu = \delta_{\mu}^0\,.
 \end{equation}
The zero-torsion condition \eqref{condition1} is sufficient but not necessary to obtain a causal non-relativistic geometry. Indeed, Frobenius' theorem states that a necessary and sufficient condition for the space-time to admit a foliation in a time flow orthogonal to Riemannian space-like leaves (and thus obey non-relativistic causality), is the so-called hypersurface orthogonality condition
\begin{equation} \label{hypsurforth}
\tau_{[\mu}\,\partial_\nu\tau_{\rho]}=0 \,,
\end{equation}
that can be equivalently written as
\begin{equation}\label{condition2}
\tau_{ab} \equiv e_a{}^\mu e_b{}^\nu \tau_{\mu\nu}=0\,,\hskip 2truecm \tau_{\mu\nu} = \partial_{[\mu}\tau_{\nu]}\,,
\end{equation}
where $e_a{}^\mu$ is the projective inverse of the spatial Vielbein $e_\mu{}^a$, with $\mu=0,1,\cdots d-1$ and $a=1,2,\cdots d-1$,  see eq.~\eqref{projinv}.
Note that in this case the time difference between two space-like leaves depends on the path between the two leaves, i.e.~there is no well-defined notion of an absolute time on which all observers agree.

The condition \eqref{condition2}, also called the twistless-torsional condition, was first encountered in the context of Lifshitz holography when studying the coupling of Newton-Cartan gravity to the Conformal Field Theory (CFT) at the boundary \cite{Christensen:2013lma}. Twistless-torsional Newton-Cartan geometry has also  been applied in studies of the Quantum Hall Effect \cite{Geracie:2014nka}. Note that it is not surprising that the more general twistless-torsional condition \eqref{condition2} was found in the context of CFTs. The zero torsion condition \eqref{condition1} is simply not allowed within a CFT since it is not invariant under space-time-dependent  dilatations $\delta\tau_\mu \sim \Lambda_D(x) \tau_\mu$. Instead, the  condition \eqref{condition2}  is invariant under space-time-dependent dilatations due to the relation $e_a{}^\mu\tau_\mu=0$, see eq.~\eqref{projinv}.

In the presence of local dilatation symmetry, one can define a conformal, i.e.~dilatation-covariant, torsion as
\begin{equation}
\tau_{\mu\nu}^C\equiv \partial_{[\mu}\tau_{\nu]} - 2b_{[\mu}\tau_{\nu]}\,,
\end{equation}
where $b_\mu$ is the gauge field of dilatations, i.e.~it transforms under dilatations as $\delta b_\mu = \partial_\mu \Lambda_D$. The twistless-torsional condition \eqref{condition2} can then also be equivalently restated as
\begin{equation}
  \tau_{\mu\nu}^C = 0 \,.
\end{equation}
Indeed, by taking the space/space projection of this equation, one obtains \eqref{condition2}:
\begin{equation}
\tau_{ab}^C \equiv e_a{}^\mu e_b{}^\nu\tau_{\mu\nu}^C = \tau_{ab} = 0\,.
\end{equation}
The space/time projection of $\tau_{\mu\nu}^C = 0$ does not lead to an extra constraint on $\tau_{\mu\nu}$, but can instead be used to solve for the spatial components of $b_\mu$:
\begin{equation}\label{solve}
\tau_{0a}^C \equiv \tau^\mu e_a{}^\nu \tau_{\mu\nu}^C = 0\hskip .5truecm \Rightarrow \hskip .5truecm b_a\equiv e_a{}^\mu b_\mu =- \tau_{0a}\,.
\end{equation}
where we used that $\tau^\mu\tau_\mu=1$, see eq.~\eqref{projinv}.

In this paper, we will be interested in considering non-relativistic geometry, both in the absence and presence of conformal symmetries, in the case of arbitrary torsion, i.e. when the zero torsion or twistless-torsional conditions no longer hold.
At first sight, it seems strange to consider the case of arbitrary torsion since causality is lost in this case.
However, in condensed matter applications, one often  considers gravity not as a dynamical theory but as background fields for determining the response of the system to a geometrical force and  for defining a non-relativistic energy and momentum flux.\footnote{This applies to the microscopic theory. Gravitational fields can
occur dynamically in an effective field theory description.} It was pointed out a long time ago in the seminal paper by Luttinger \cite{Luttinger:1964zz} that to describe thermal transport in a resistive medium one needs to consider an auxiliary  gravitational field $\psi(x)$ that couples to the energy and is defined by
\cite{Gromov:2014vla}
\begin{equation}
\tau_\mu = e^{\psi(x)}\delta_{\mu}^0 \,,
\end{equation}
corresponding to the case of twistless torsion. Later, it was pointed out that, for describing other properties as well, one also needs to
introduce the other components of $\tau_\mu$  that couple to the energy current. This leads to a non-relativistic energy-momentum tensor with no
restrictions and an un-restricted $\tau_\mu$
describing arbitrary torsion \cite{Gromov:2014vla}. For other applications of torsion in condensed matter, see \cite{Geracie:2014mta, Geracie:2016dpu}.\,\footnote{In \cite{Geracie:2016dpu} non-zero expressions for the spatial torsion, i.e.~the curl of the spatial Vielbein,
and for the curl of the central charge gauge field are considered as well. We will not consider this more general situation here.} To avoid confusion, we will reserve the word `geometry' if we only consider the background fields and their symmetries whereas we will talk about `gravity'  if
these background fields satisfy dynamical equations of motion.

In this paper, we will construct by two complementary techniques, gauging and null-reduction, the extension of NC geometry
and its non-relativistic conformal extension, Schr\"odinger geometry with dynamical exponent $z=2$, to the case of arbitrary torsion, i.e.~$\tau_{\mu\nu}\ne 0$ for NC geometry and $\tau_{ab}\ne0$ for Schr\"odinger geometry, see Table \ref{NCtorsiontab}. Furthermore, applying a different technique thereby making use of the obtained results on Schr\"odinger geometry with arbitrary torsion, we will construct the extension of NC gravity to the case of arbitrary torsion, in three space-time dimensions.
Note that in the conformal case we will always impose that $\tau_{0a}^C=0$, i.e.~the minimal torsion case is twistless-torsional, in agreement with the fact that the zero torsion condition is incompatible with dilatation symmetry. As explained above, $\tau_{0a}^C = 0$ does not lead to a constraint on $\tau_{\mu\nu}$. Rather it is a so-called conventional constraint, that can be used to solve for $b_a$, see eq.~\eqref{solve}. For earlier discussions of Newton-Cartan geometry with torsion and null-reductions, see \cite{Duval:1984cj,Julia:1994bs,Jensen:2014aia,Bekaert:2014bwa,Festuccia:2016awg}.

This paper is organized as follows. In section 2 we will apply the gauging technique to the Bargmann algebra in $d$ space-time dimensions. In particular, we will  construct the transformation rules of the independent fields and the expressions of the dependent spin-connections  of NC geometry for the case of arbitrary torsion.
In section  3 we derive the same results
from an off-shell, meaning we do not reduce the equations of motion, null-reduction of General Relativity in $d+1$ space-time dimensions.
 We point out that performing a null-reduction of the equations of motion as well we obtain the equations of motion of  NC gravity with {\sl zero}  torsion
 thereby reproducing the result of \cite{Julia:1994bs}.  We point out that the zero torsion condition is related to the invariance under central charge transformations that necessarily follows from the null-reduction. To obtain NC gravity with arbitrary torsion, we will first in the next two sections repeat the calculations of sections 2 and 3 but now for the minimal conformal extension of the Bargmann algebra, i.e.~the Schr\"odinger algebra, and for conformal gravity except
  that we do not consider the equations of motion in this case. To be precise, in section 4 we will gauge the $z=2$ Schr\"odinger algebra and obtain the transformation rules of $z=2$ Schr\"odinger geometry for arbitrary torsion together with the expressions of the dependent gauge fields. Next, in section 5, we obtain the same results by performing a null-reduction of conformal gravity in $d+1$ space-time dimensions. In section 6, we use these results to construct three-dimensional NC gravity with arbitrary torsion by starting from a $z=2$ Schr\"odinger Field Theory (SFT) for a complex compensating scalar, coupling it to the Schr\"odinger geometry with arbitrary torsion we constructed in sections 4 and 5 and gauge-fixing the dilatations and central charge transformations.  We give our comments in the Conclusions.

\begin{table}[h!]
\caption{Newton-Cartan and Schr\"odinger geometry with torsion.}\label{NCtorsiontab}
\begin{center}
\begin{tabular}{|c|c||c|c|}
\hline
&&&\\[-.4truecm]
 geometric constraint& Newton-Cartan& geometric constraint&Schr\"odinger \\ [.1truecm] \hline
 &&&\\[-.4truecm]
 $\tau_{0a}\ne 0\,, \tau_{ab}\ne 0$&arbitrary torsion&$\tau_{ab}\ne 0$&arbitrary torsion\\[.1truecm]
 $\tau_{0a}\ne 0\,, \tau_{ab}= 0$&twistless-torsional&$\tau_{ab}=0$&
 twistless-torsional\\[.1truecm]
 $\tau_{0a}=0\,, \tau_{ab}= 0$&zero torsion&--&--\\
 \hline
\end{tabular}
\end{center}
\end{table}

\section{Gauging the Bargmann Algebra with Arbitrary Torsion}
\label{appA}

Our starting point is the $d$-dimensional Bargmann algebra whose non-zero commutators are given by
\begin{align}
[J_{ab},J_{cd}] & = 4 \delta_{[a[c}J_{d]b]}\,, \hskip 1truecm
[J_{ab},P_c]  = -2\delta_{c[a}P_{b]}\,, \nonumber\\[.1truecm]
[J_{ab}, G_c] & = -2 \delta_{c[a}G_{b]}\,,\hskip 1truecm
[G_a, H]  = - P_a\,, \nonumber\\[.1truecm]
[G_a,P_b] & = -\delta_{ab}M \,,\label{Bargmannalgebra}
\end{align}
where
\begin{equation}\label{generators}
\{H\,, P_a\,, J_{ab}\,, G_a\,, M\}
 \end{equation}
 are the generators corresponding to time translations, spatial translations, spatial rotations, Galilean boosts and central charge transformations, respectively. Note that the $M$-generator has the dimension of a mass and that for $M=0$  the Bargmann algebra reduces to the Galilei algebra. The gauging of the Bargmann algebra for zero torsion has been considered in \cite{Andringa:2010it}. In this section we will extend this gauging to the case of arbitrary torsion, see also
\cite{Festuccia:2016awg}.

 The gauge fields corresponding to the generators \eqref{generators} are given by
 \begin{equation}
 \{\tau_\mu\,, e_\mu{}^a\,, \omega_\mu{}^{ab}\,, \omega_\mu{}^a\,, m_\mu\},
 \end{equation}
 respectively. Under general coordinate transformations, they transform as covariant vectors. Under the spatial rotations, Galilean boosts and central charge transformations, with parameters $\{\lambda^a{}_b\,,\lambda^a\,, \sigma\}$, respectively, the gauge fields $\{\tau_\mu\,, e_\mu{}^a\,, m_\mu\}$
 that will remain independent, see below, transform according to the structure constants of the Bargmann algebra,
 i.e.:
 \bea\label{transfB}
\d \tau_{\m}&=&0~,\nonumber
\\[.1truecm]
\d e_{\m}{}^{a}&=&\l^a{}_{b}e_{\m}{}^b + \l^a\tau_{\m}~,
\\[.1truecm]
\d m_{\m}&=&\partial_{\m}\s+\l^ae_{\m a}~.\nonumber
\eea
These independent fields and their transformation rules then define NC geometry in the presence of arbitrary torsion, i.e.~$\tau_{\mu\nu}\ne 0$.

Now that we have arbitrary torsion,  we can modify the usual conventional constraints that can be solved for the spin-connection fields
$\omega_\mu{}^{ab}$ and $\omega_\mu{}^a$ such that these spin-connections receive torsion contributions. We choose the following conventional constraints
that are justified by the null-reduction of General Relativity that we will perform in the next section:
\bea\label{constraints1}
&&R_{\mu\nu}(P^a) +2 \tau^a{}_{[\mu}m_{\nu]} = 0 \,,\\[.2truecm]
&&R_{\mu\nu}(M) - 2 \tau_{0[\mu} m_{\nu]}=0\,,\label{constraints2}
\eea
with $\tau_{0\nu}\equiv \tau^\mu \tau_{\mu\nu}\,,\tau_{a\nu}\equiv e_a{}^\mu \tau_{\mu\nu}$ and with the curvatures $R_{\mu\nu}(P^a)$ and $R_{\mu\nu}(M)$ given by expressions that follow from the structure constants of the Bargmann algebra:
\begin{align}
  \label{curvatures}
  R_{\mu\nu}(P^a) &= 2 \partial_{[\mu} e_{\nu]}{}^a - 2 \omega_{[\mu}{}^{ab} e_{\nu]b} - 2 \omega_{[\mu}{}^a \tau_{\nu]} \,, \nonumber \\[.1truecm]
  R_{\mu\nu}(M) &= 2 \partial_{[\mu} m_{\nu]} - 2 \omega_{[\mu}{}^a e_{\nu]a} \,.
\end{align}

Explicitly, the  expressions for the torsionful spin-connections that follow from the constraints \eqref{constraints1} and \eqref{constraints2} are
given by\,\footnote{Note that from now on the spin-connections are dependent fields.
In cases, when confusion could arise, we will indicate the explicit dependence.}
\begin{alignat}{2}
  \label{eq:omegareduxcurvedsimple}
  & \omega_\mu{}^{ab}(\tau,e,m) = \mathring{\omega}_\mu{}^{ab}(\tau,e,m) - m_\mu \tau^{ab}\,,\nonumber \\[.2truecm]
  & \omega_\mu{}^{a} (\tau,e,m) =  \mathring{\omega}_\mu{}^a (\tau,e,m)+  m_\mu \tau_0{}^a \,,
 \end{alignat}
 where the space/space  and space/time components of the torsion are given by
\be
  \tau_{ab} = e^{\mu}{}_a e^{\nu}{}_b \partial_{[\mu} \tau_{\nu]} \,, \qquad
  \tau_{0a} = \tau^\mu e^{\nu}{}_a \partial_{[\mu} \tau_{\nu]}
\ee
and where $\mathring{\omega}_{\m}{}{}^{ab}(\tau,e,m)$ and $\mathring{\omega}_{\m}{}^{a}(\tau,e,m)$ are the torsion-free Newton-Cartan spin-connections given by
\bea
\label{ncsc1} \mathring{\omega}_{\mu}{}^{ab}(\tau,e,m)&=&e_{\mu c} e^{\rho a} e^{\sigma b} \partial_{[\rho} e_{\sigma]}{}^c
    - e^{\nu a} \partial_{[\mu} e_{\nu]}{}^b + e^{\nu b} \partial_{[\mu} e_{\nu]}{} ^a - \tau_\mu e^{\rho a} e^{\sigma b}
      \partial_{[\rho} m_{\sigma]} ~,\\[.2truecm]
\label{ncsc2} \mathring{\omega}_\mu{}^{a}(\tau,e,m) &=&  \tau^\nu \partial_{[\mu} e_{\nu]}{}^a +  e_\mu{}^c e^{\rho a}
    \tau^\sigma \partial_{[\rho} e_{\sigma]c} +  e^{\nu a} \partial_{[\mu} m_{\nu]} +  \tau_\mu \tau^\rho e^{\sigma a}
      \partial_{[\rho} m_{\sigma]}~.
\eea

The expressions  for $\mathring{\omega}_{\mu}{}^{ab}$ and $\mathring{\omega}_\mu{}^{a}$ are the solutions of the constraints \eqref{constraints1} and \eqref{constraints2} for zero torsion, i.e.~$\tau_{\mu\nu}=0$. Note that the solutions
\eqref{ncsc1} and \eqref{ncsc2} contain the fields $\tau^\mu$ and $e^\mu{}_a$ that are defined by the following projective invertibility relations
\bea\label{projinv}
e^{\m}{}_{a}e_{\nu}{}^{a}&=&\d^{\m}_{\nu}-\tau^{\m}\tau_{\nu}~,\qquad
e^{\m}{}_{a}e_{\m}{}^{b}=\d^a_b~,\nonumber
\\[.2truecm]
\tau^{\m}\tau_{\m}&=&1~,\qquad
e^{\m}{}_{a}\tau_{\m}=0~,\qquad
\tau^{\m}e_{\m}{}^{a}=0~.
\eea

It is important to note that the dependent torsion-free spin-connections $ \mathring{\omega}_\mu{}^{ab}(\tau,e,m)$ and $\mathring{\omega}_\mu{}^{a} (\tau,e,m)$, due to the arbitrary torsion, no longer transform according to the Bargmann algebra. In particular, from eqs.~\eqref{ncsc1} and \eqref{ncsc2} it follows that  their transformation rules under Galilean boosts contain extra torsion terms given by
\begin{align}
\label{omegaB}
\Delta\mathring{\omega}_\mu{}^{ab} &=  \lambda^c e_{\mu c}\tau^{ab} + 2\lambda^{[a}e^{|\rho|b]}\tau_{\mu\rho}\,, \nonumber\\[.1truecm]
\Delta\mathring{\omega}_\mu{}^{a} &= -\lambda^a e_\mu{}^b\tau_{0b} -\lambda^b e_{\mu b}\tau_0{}^a\,.
\end{align}
Correspondingly, the curvatures corresponding to these spin-connections that transform covariantly under Galilean boosts contain extra torsion contributions and are given by
\begin{align}
  \label{eq:deflowercurvs}
R_{\mu\nu}(J^{ab}) &= 2 \partial_{[\mu} \mathring{\omega}_{\nu]}^{ab} - 2 \mathring{\omega}_{[\mu}{}^{ac} \mathring{\omega}_{\nu]c}{}^b- 2 \mathring{\omega}_{[\mu}{}^c e_{\nu]c} \tau^{ab} - 4 \mathring{\omega}_{[\mu}{}^{[a} e^{|\rho|b]} \tau_{\nu]\rho} \,,\nonumber \\[.1truecm]
R_{\mu\nu}(G^a) &= 2 \partial_{[\mu} \mathring{\omega}_{\nu]}{}^a - 2 \mathring{\omega}_{[\mu}{}^{ab} \mathring{\omega}_{\nu] b} + 2 \mathring{\omega}_{[\mu}{}^a e_{\nu]}{}^b \tau_{0b}+ 2 \mathring{\omega}_{[\mu}{}^b e_{\nu] b} \tau_0{}^{a} \,.
\end{align}
These are the curvatures that naturally appear in the next section when we perform a null-reduction of the equations of motion of General Relativity,  see eq.~\eqref{flatRiccicomp}.
Note that there is an arbitrariness in the definition of these curvatures in the sense that one can always move around torsion terms in or outside the spin-connections.
In that sense the above curvatures are defined modulo $D\tau$ and $\tau^2$ terms. The specific definition we use naturally follows from the null-reduction in the next section.

\section{The Null-reduction of General Relativity}
\label{sec2}

In this section we re-obtain the results on NC geometry with arbitrary torsion obtained in the previous section by performing  a dimensional reduction of  General Relativity (GR) from $d+1$ to $d$ space-time dimensions along a null-direction \cite{Duval:1984cj,Julia:1994bs}. We show that in this way one obtains the same transformation rules and the same expressions for the dependent spin-connections as before. Next, we point out that, after going on-shell, the equations of motion reduce to those of NC gravity with {\sl zero}  torsion \cite{Duval:1984cj,Julia:1994bs}.

Our starting point is General Relativity in $d+1$ dimensions in the second order formalism, where the single independent field
is the Vielbein $\hat{e}_M{}^{A}$. Here and in the following, hatted fields are $(d+1)$-dimensional and unhatted ones will denote
$d$-dimensional fields after dimensional reduction. Furthermore, capital indices take $d+1$ values, with $M$ being a
curved  and $A$ a flat  index. The Einstein-Hilbert action in $d+1$ space-time dimensions is given by
\be
S_{\text{GR}}^{(d+1)}=-\frac 1{2\kappa}\int \dd^{d+1}x \, \hat{e} \,
  \hat{e}^{M}{}_{A}\hat{e}^{N}{}_{B}\hat R_{MN}{}{}^{AB}\left(\hat\omega(\hat e)\right)~,
\ee
where $\kappa$ is the gravitational coupling constant and $\hat e$ is the determinant of the Vielbein. The inverse Vielbein satisfies the usual relations
\be \label{inverses}
\hat e^M{}_{A}\hat e_M{}^{B}=\d^B_A~, \quad \hat{e}^M{}_{A}\hat e_N{}^{A}=\d^M_N~.
\ee
The spin-connection is a dependent field, given in terms of the vielbein as
\be \label{spinconn}
\hat{\omega}_M{}^{BA}(\hat e)=2\hat{e}^{N[A}\partial_{[M}\hat e_{N]}{}^{B]}-\hat e^{N[A}\hat e^{B]P}\hat e_{MC}\partial_{N}\hat e_P{}^{C}~,
\ee
while the curvature tensor is given by
\be \label{curv}
\hat{R}_{MN}{}^{AB}\left(\hat\omega(\hat e)\right) = 2 \partial_{[M} \hat{\omega}_{N]}{}^{AB} - 2 \hat{\omega}_{[M}{}^{AC} \hat{\omega}_{N]C}{}^B \,.
\ee
Under infinitesinal general coordinate transformations, with parameter $\zeta^M$ and local Lorentz transformations, with parameter $\lambda^A{}_B$, the Vielbein transforms as
\be \label{vielbeintransformation}
\d\hat{e}_M{}^A=\zeta^N\partial_N\hat e_M{}^{A}+\partial_M\zeta^N\hat e_{N}{}^{A} + \lambda^A{}_B\hat e_M{}^B~.
\ee

In order to dimensionally reduce the transformation rules  along a null-direction, we assume the existence of a null Killing vector $\xi=\xi^M\partial_M$
for the metric $\hat{g}_{MN}\equiv {\hat e}_M{}^A{\hat e}_N{}^B\eta_{AB}$, i.e.
\be
{\cal L}_{\xi}\hat{g}_{MN}=0 \quad \text{and} \quad \xi^2=0~.
\ee
Without loss of generality, we may choose adapted coordinates $x^M=\{x^{\m},\u\}$, with $\m$ taking $d$ values,
and take the Killing vector to be $\xi=\xi^{v}\partial_{\u}$. Then the Killing equation implies that the metric is $\u$-independent,
i.e.~$\partial_{\u}\hat{g}_{MN}=0$,
while the null condition implies the following constraint on the metric:\,\footnote{Due to this constraint, we are not allowed to perform the
null-reduction in the action but only in the transformation rules and equations of motion \cite{Julia:1994bs}.}
\begin{equation}
\hat {g}_{vv}=0\,.
\end{equation}
A  suitable reduction Ansatz for the
Vielbein should be consistent with this constraint on the metric. Such an Ansatz was discussed in  \cite{Julia:1994bs}, and we repeat it below
in a formalism suited to our purposes.

First, we split the $(d+1)$-dimensional tangent space indices as $A=\{a,+,-\}$, where the index $a$ is purely spatial and takes
$d-1$ values, while $\pm$ denote null directions. Then the Minkowski metric components are $\eta_{ab}=\d_{ab}$ and $\eta_{+-}=1$.
The reduction Ansatz is specified upon choosing the inverse Vielbein $\hat{e}^M{}_{+}$ to be proportional to the null
Killing vector $\xi=\xi^v\partial_v$. A consistent parametrization is
\begin{equation}
  \label{eq:invvielbansatz}
  \hat e^M{}_A = \bordermatrix{& \mu & v \cr\\[.2truecm]
a & e^{\mu}{}_a & e^\mu{}_a m_\mu \cr\\[.2truecm]
- & S\tau^\mu & S \tau^\mu m_\mu \cr
+ & 0 & S^{-1} }\,.
\end{equation}
The scalar $S$ is a compensating one and can be gauge-fixed as we will see shortly.

Given the expression \eqref{eq:invvielbansatz} for the inverse Vielbein, the Vielbein itself is  given by
\begin{equation}
  \label{eq:vielbeinansatz}
  \hat e_M{}^A = \bordermatrix{& a & - & + \cr\\[.2truecm]
\mu & e_\mu{}^a & S^{-1} \tau_\mu & -S m_\mu \cr
v & 0 & 0 & S} \,.
\end{equation}
To avoid confusion, recall that the index $a$ takes one value less than the index $\mu$; thus the above matrices are both
square although in block form this is not manifest.

Note that the Ansatz \eqref{eq:vielbeinansatz} has two zeros. The zero in the second column, ${\hat e}_v{}^-=0$, is due to the existence of the null Killing vector $\xi=\xi^v\partial_v$:
\be
\xi^2 = \xi^v\xi^v {\hat g}_{vv}= 0 \quad \Rightarrow \quad
\hat{g}_{\u\u}=\hat{e}_{\u}{}^A\hat{e}_{\u}{}^B\eta_{AB} =0\quad \Rightarrow \quad
\hat{e}_{\u}{}^-=0~.
\ee
On the other hand, the zero in the first column, ${\hat e}_v{}^a=0$, implies that the Lorentz transformations with parameters $\lambda^a{}_+$ are gauge-fixed. We are thus left over with $\lambda^a{}_b$, $\lambda^a{}_-$, that we will call $\lambda^a \equiv \lambda^a{}_-$, and $\lambda^+{}_+=-\lambda^-{}_-$, that we will call $\lambda$. The latter can be gauge-fixed by imposing $S=1$. For some purposes, especially when we discuss the conformal case, it is convenient to only perform this gauge-fixing at a later stage, so we will momentarily keep $S$.

A simple computation reveals that
the invertibility relations \eqref{inverses}, after substitution of the reduction Ansatz, precisely reproduce the projective invertibility relations \eqref{projinv} encountered
when gauging the Bargmann algebra provided we identify $\{\tau_\mu\,, e_\mu{}^a\}$ as the timelike and spatial Vielbein of NC gravity, respectively.

Starting from the transformation rule \eqref{vielbeintransformation} of the $(d+1)$-dimensional Vielbein, we derive the
following transformations of the lower-dimensional fields:
\bea
\d \tau_{\m}&=&0~,
\\[.1truecm]
\d e_{\m}{}^{a}&=&\l^a{}_{b}e_{\m}{}^b + S^{-1}\l^a\tau_{\m}~,
\\[.1truecm]
\d m_{\m}&=&-\partial_{\mu}\zeta^{v}-S^{-1}\l_ae_{\m}{}^a~,\\[.1truecm]
\delta S &=& \lambda S\,,
\eea
where $\zeta^v$ denotes the component of the parameter of $(d+1)$-dimensional diffeomorphisms, along the compact $v$-direction.
Next, fixing the Lorentz transformations with parameter $\lambda$ by setting $S=1$ and defining $\sigma:=-\zeta^{v}$
we precisely obtain the transformation rules \eqref{transfB} of Newton-Cartan geometry in $d$ dimensions provided we identify $m_\mu$ as the central charge gauge field
 associated to the  central charge generator of the Bargmann algebra. Note that we have not imposed any constraint on the torsion, i.e.~$\tau_{\mu\nu} =\partial_{[\mu}\tau_{\nu]}  \ne 0$.

We next consider the null-reduction of the spin-connection given in \eqref{spinconn}.
Inserting the Vielbein Ansatz \eqref{eq:vielbeinansatz} with $S=1$ into \eqref{spinconn} we obtain the following expressions for the different components:
\begin{alignat}{2}
  \label{eq:omegareduxcurvedsimple2}
  & \hat{\omega}_\mu{}^{ab}(\hat e) \equiv  \omega_\mu{}^{ab}(\tau,e,m) = \mathring{\omega}_\mu{}^{ab}(e,\tau,m) - m_\mu \tau^{ab} \,,
   \nonumber \\[.2truecm]
  & \hat{\omega}_\mu{}^{a+} (\hat e) \equiv  \omega_\mu{}^a(\tau,e,m) = \mathring{\omega}_\mu{}^a (e,\tau,m)+  m_\mu \tau_0{}^a \,, \nonumber \\[.2truecm]
&\hat{\omega}_v{}^{ab} (\hat e) = \tau^{ab} \,, \qquad & &    \hat{\omega}_v{}^{a+}(\hat e)  = - \tau_0{}^a \,, \nonumber \\[.2truecm]
  & \hat{\omega}_\mu{}^{a-}(\hat e)  = -   \tau_\mu \tau_0{}^a - e_\mu{}^b \tau_b{}^a  \,, \qquad & & \hat{\omega}_v{}^{a-}(\hat e)  = 0 \,, \nonumber \\[.2truecm]
  & \hat{\omega}_\mu{}^{-+} (\hat e) = -e_\mu{}^b \tau_{0b} \,, \qquad & & \hat{\omega}_v{}^{-+} (\hat e) = 0 \,,
\end{alignat}
where $\mathring{\omega}_{\m}{}{}^{ab}(e,\tau,m)$ and $\mathring{\omega}_{\m}{}^{a}(e,\tau,m)$ are the torsion-free  Newton-Cartan spin-connections given in
eqs.~\eqref{ncsc1} and \eqref{ncsc2}. Note that the first two lines precisely reproduce the expressions for the torsionful spin-connections of NC gravity given in
eqs.~\eqref{eq:omegareduxcurvedsimple} of the previous section.

At this point, we have re-produced using the complementary null-reduction technique the results on NC geometry with arbitrary torsion obtained in the  previous section. To calculate the equations of motion after null-reduction, we first need to calculate the components of the higher-dimensional Ricci tensor with flat indices:
\be\label{Ricci}
\hat{R}_{AB}\left(\hat\omega(\hat e)\right)=\hat{e}^M{}_{C}\hat{e}^N{}_{A}\hat{R}_{MN}{}^C{}_B\left(\hat\omega(\hat e)\right)\,.
\ee
Substituting the reduction Ansatz \eqref{eq:vielbeinansatz}, with $S=1$, into \eqref{Ricci} we find the following expressions for the Ricci tensor components:
\begin{align}
  \label{flatRiccicomp}
  \hat{R}_{++} &= - \tau^{ab} \tau_{ab} \,, \nonumber \\[.1truecm]
  \hat{R}_{+-} &= -D_a \tau_0{}^{a} + 2 \tau_0{}^{a} \tau_{0a} \,, \nonumber \\[.1truecm]
  \hat{R}_{--} &= - R_{0a}(G^a)  \,, \nonumber \\[.1truecm]
  \hat{R}_{+a} &=  D_b \tau^b{}_a - 2  \tau_0{}^{b} \tau_{ba} \,, \nonumber \\[.1truecm]
  \hat{R}_{-a} &= - R_{0b}(J^b{}_a) -  D_0 \tau_{0a} \,, \nonumber \\[.1truecm]
  \hat{R}_{ab} &= R_{ca}(J^c{}_b) - 2 D_{a} \tau_{0 b} +  D_0 \tau_{ab} + 2 \tau_{0a} \tau_{0b} \,,
\end{align}
where the lower-dimensional curvatures $R(J)$ and $R(G)$ are defined in eq.~\eqref{eq:deflowercurvs} and where the covariant derivatives on $\tau_0{}^{a}$ and  $\tau^{ab}$ are given by
\begin{align}
  \label{eq:defcovdertorsion}
  D_\mu \tau_0{}^{a} &= \partial_\mu \tau_0{}^{a} - \mathring{\omega}_\mu{}^{ab} \tau_{0b} + \mathring{\omega}_\mu{}^b \tau_b{}^a \,, \nonumber \\[.1truecm]
  D_\mu \tau^{ab} &= \partial_\mu \tau^{ab} - \mathring{\omega}_\mu{}^a{}_c \tau^{cb} - \mathring{\omega}_\mu{}^b{}_c \tau^{ac} \,.
\end{align}
Using the Bianchi identity for $\tau_{\mu\nu}$ in the form
\begin{equation}
  D_0 \tau_{ab} = D_a \tau_{0b} - D_b \tau_{0a} \,,
\end{equation}
we can rewrite the Ricci tensor components $\hat{R}_{ab}$ in a manifestly symmetric form as follows:
\begin{equation}
  \hat{R}_{ab} = R_{ca}(J^c{}_b) - 2 D_{(a} \tau_{|0|b)} + 2 \tau_{0a} \tau_{0b} \,.
\end{equation}

We first consider  the Ricci tensor components that contain the curvatures $R(J)$ and/or $R(G)$. They lead to the following set of equations of motion:
\begin{align}\label{eom}
R_{0a}(G^a)&=0  \,, \qquad
 R_{c\bar{a}}(J^c{}_b) - 2 D_{(\bar{a}} \tau_{|0| b)} + 2 \tau_{0\bar{a}} \tau_{0b} =0\,,
\end{align}
where in the last equation we collected  two field equations into one by using an index $\bar{a}=(a,0)$. At first sight, it looks like this first set of equations of motion defines NC gravity with arbitrary torsion. However, the other set of equations, obtained by putting $\hat{R}_{++}$, $\hat{R}_{+-}$ and $\hat{R}_{+a}$ to zero, cannot be ignored and they constrain the torsion. For instance, the equation $\hat {R}_{++}=0$ implies $\tau_{ab}=0$ while the equation  $\hat {R}_{+-}=0$ implies, with a proper choice of boundary conditions,  $\tau_{0a}=0$.
Since the first set of equations of motion transforms to the second one under Galilean boosts, it is not consistent to  leave out the second set of equations of motion in the hope of obtaining NC equations of motion with arbitrary torsion.
Together, they imply zero torsion and, after substituting this back into \eqref{eom}, one obtains the equations of motion corresponding to NC gravity with zero torsion \cite{Julia:1994bs}.
\begin{align}\label{eomzerotorsion}
R_{0a}(G^a)&=0  \,, \qquad
R_{c0}{}^c{}_b(J)  =0\,, \, \qquad     R_{ca}{}^c{}_b(J)  =0\,.
\end{align}

\section{Gauging the $z=2$ Schr\"odinger Algebra  with Arbitrary Torsion}
\label{appB}

In this section we extend the gauging of the so-called $z=2$ Schr\"odinger algebra with twistless torsion as performed in \cite{Bergshoeff:2014uea} to the case of arbitrary torsion, i.e.~$\tau_{ab}\ne 0$. Our starting point  is the $z=2$ Schr\"odinger algebra which is the minimal conformal extension, with dynamical exponent $z=2$,  of the $d$-dimensional Bargmann algebra whose commutation relations were given in eq.~\eqref{Bargmannalgebra}. To this end we add the additional generators
$D$ and $K$ corresponding to dilatations and special conformal transformations with  gauge fields $b_\mu$ and $f_\mu$, respectively. The additional non-zero commutation relations with respect to the Bargmann algebra are given by
\begin{alignat}{2} \label{Schroedingeralgebra}
\left[D,H\right] &= -2H\,,& \qquad \left[H,K\right] &=  D\,,\nonumber \\[.1truecm]
\left[D,K\right] &= 2K\,,& \left[K,P_a\right] &= -G_a\,,\nonumber \\[.1truecm]
\left[D,P_a\right] &= -P_a\,, & \left[D,G_a\right] &=  G_a\,.\nonumber
\end{alignat}

This leads us to the following complete set of covariant one-form gauge fields:
\be
\{e_{\mu}{}^a,\tau_{\m},\omega_{\m}{}^{ab},\omega_{\mu}{}^a,b_{\m},f_{\m},m_{\m}\}\,.
\ee
Only the subset $\{\tau_\mu, e_\mu{}^a, m_\mu,  b_0\}$, with $b_0\equiv \tau^\mu b_\mu$, will remain independent gauge fields. Following the structure constants of the Schr\"odinger algebra these independent gauge fields transform under the Bargmann symmetries and the additional dilatations, with parameter $\lambda_D$, and special conformal transformations, with parameter $\lambda_K$, as follows:
\begin{align}\label{transfSchr}
\delta\tau_\mu & =  2\lambda_D\tau_\mu\,,\nonumber \\[.1truecm]
\delta e_\mu{}^a & =  \lambda^a{}_b e_\mu{}^b+\lambda^a\tau_\mu+\lambda_De_\mu{}^a\,,\nonumber \\[.1truecm]
\delta m_\mu & =  \partial_\mu\sigma+\lambda^a e_{\mu a}\,,\nonumber\\[.1truecm]
\delta  b_0 & =   \partial_0\lambda_D+\lambda_K - \lambda^a e_a{}^\mu b_\mu \,.
\end{align}

We now impose the following first set of  conventional curvature constraints:\,\footnote{We indicate the Schr\"odinger curvatures with a script ${\cal R}$. Note that, in contrast to \cite{Bergshoeff:2014uea}, we do not impose that ${\cal R}_{ab}(H)=0$, i.e.~we have
arbitrary torsion: $\tfrac{1}{2}{\cal R}_{ab}(H) = \tau^C_{ab} = \tau_{ab}\ne 0$. We have chosen the second conventional constraint such that it gives the same torsionful rotational spin-connection that follows from the null-reduction that we will perform in the next section. }
\begin{eqnarray}\label{conv2}
 {\cal R}_{0a}(H) = 0\,,\nonumber\\[.1truecm]
 {\cal R}_{\mu\nu}{}^a(P) + 2 \tau^{C\,a}{}_{[\mu}m_{\nu]} = 0\,,\nonumber\\[.1truecm]
{\cal R}_{\mu\nu}(M) = 0\,.
\end{eqnarray}
We have used here the following curvatures whose expressions follow from the structure constants of the Schr\"odinger algebra:
\begin{align} \label{curvatureszis2}
{\cal R}_{\mu\nu}(H) & =  2\partial_{[\mu}\tau_{\nu]}-4b_{[\mu}\tau_{\nu]}\,,\nonumber \\[.1truecm]
{\cal R}_{\mu\nu}{}^a(P) & =  2\partial_{[\mu}e_{\nu]}{}^a-2\omega_{[\mu}{}^{ab}e_{\nu]b}-2\omega_{[\mu}{}^a\tau_{\nu]}-2b_{[\mu}e_{\nu]}{}^a\,, \nonumber \\[.1truecm]
{\cal R}_{\mu\nu}(M) & =  2\partial_{[\mu}m_{\nu]}-2\omega_{[\mu}{}^ae_{\nu]a}\,.
\end{align}

The conventional constraints  \eqref{conv2} allow us to solve for the spatial components of  $b_\mu$ and of the spin-connection fields
$\omega_\mu{}^{ab}$ and $\omega_\mu{}^a$ as follows\,{\footnote{The only notational difference with respect to \cite{Bergshoeff:2014uea} is that in that paper the projective inverse of
$\tau_{\mu}$ is denoted as $\u^{\mu}$ and it is related to the one we use here by $\u^{\m}=-\tau^{\mu}$.}}
\bea
b_a&=&-\tau_{0a}~,\\[.1truecm]
\label{schrspin1}
\omega_{\mu}{}^{ab}(e,\tau,m,b)&=&\mathring{\omega}_{\mu}{}^{ab}(e,\tau,m,b) -m_\mu\tau^{ab}~,\\[.1truecm]
\label{schrspin2}
\omega_{\mu}{}^{a}(e,\tau,m,b)&=&\mathring{\omega}_{\mu}{}^a(e,\tau,m,b)\,,
\eea
where the torsionless Schr\"odinger spin-connections, i.e.~the part with $\tau_{ab}=0$, are related to the torsionless Newton-Cartan spin-connections defined in eqs.~\eqref{ncsc1} and \eqref{ncsc2} as follows\,{\footnote{Note that we commit some abuse of notation here, by using the same symbol $\mathring{\omega}$ for $\mathring{\omega}_\mu{}^{ab}(e,\tau,m)$, $\mathring{\omega}_\mu{}^a(e,\tau,m)$ and $\mathring{\omega}_\mu{}^{ab}(e,\tau,m,b)$, $\mathring{\omega}_\mu{}^a(e,\tau,m,b)$. For the rest of this paper, $\mathring{\omega}_\mu{}^{ab}$ and $\mathring{\omega}_\mu{}^a$ will always refer to $\mathring{\omega}_\mu{}^{ab}(e,\tau,m,b)$, $\mathring{\omega}_\mu{}^a(e,\tau,m,b)$.}}:
\bea
\label{schr1}
\mathring{\omega}_{\mu}{}^{ab}(e,\tau,m,b)&=& \mathring{\omega}_{\mu}{}^{ab}(e,\tau,m)+2e_{\mu}{}^{[a}b^{b]}~,\\[.1truecm]
\label{schr2}
\mathring{\omega}_{\mu}{}^a(e,\tau,m,b)&=& \mathring{\omega}_{\mu}{}^a(e,\tau,m)+e_{\mu}{}^ab_{0}\,.
\eea

In a second step, to solve for the gauge field $f_\mu$, we impose the following second set of conventional constraints:
\begin{eqnarray}
{\cal R}_{a0}(D) +{\cal R}_{ab}{}^b(G)-\sfrac 1{2d} m_a{\cal R}_{bc}{}^{bc}(J) &=&0 \,,\\[.1truecm]
{\cal R}_{0a}{}^a(G) -\sfrac 1{2d}m_0 {\cal R}_{ab}{}^{ab}(J) &=&0\,,
\end{eqnarray}
where the expressions for the  curvatures are given by
\begin{align} \label{curvatureszis3}
{\cal R}_{\mu\nu}{}^{ab}(J) & =  2\partial_{[\mu}\mathring{\omega}_{\nu]}{}^{ab}-2\mathring{\omega}_{[\mu}{}^{ca}\mathring{\omega}_{\nu]}{}^{b}{}_c
- 2 \mathring{\omega}_{[\mu}{}^c e_{\nu]c} \tau^{ab} + 4 \mathring{\omega}_{[\mu}{}^{[a} \tau^{b]c}e_{\nu]c}-4\mathring{\omega}_{[\mu}{}^ce_{\nu]}{}^{[a}\tau_c{}^{b]}\,,\nonumber \\[.1truecm]
{\cal R}_{\mu\nu}{}^a(G) & =  2\partial_{[\mu}\mathring\omega_{\nu]}{}^a+2\mathring\omega_{[\mu}{}^b\mathring\omega_{\nu]}{}^a{}_b-2\mathring\omega_{[\mu}{}^ab_{\nu]}-2f_{[\mu}e_{\nu]}{}^{a}\,,\nonumber \\[.1truecm]
{\cal R}_{\mu\nu}(D) & =  2\partial_{[\mu}b_{\nu]}-2f_{[\mu}\tau_{\nu]}+2\mathring{\omega}_{[\mu}{}^be_{\nu]}{}^a\tau_{ab}\,.
\end{align}
Note that these curvatures, save the one corresponding to $G^a$, contain extra torsion contributions that render them covariant under Galilean boosts.
This second set of conventional constraints is chosen such that it precisely reproduces the same expression for $f_\mu$ that we will derive in the next section by a null-reduction of conformal gravity:
\begin{eqnarray}\label{fsolved}
&&f_a= \frac 1{d-1} {\cal R}^\prime_{a0}(D) +\frac 1{d-1} {\cal R}'_{ab}{}^b(G)-\frac 1{2d(d-1)} m_a{\cal R}_{bc}{}^{bc}(J) \,,\\[.1truecm]
&& f_0 = \frac{1}{d-1}{\cal R}^\prime_{0a}{}^a(G) -\frac 1{2d(d-1)}m_{0} {\cal R}_{ab}{}^{ab}(J)\,.
\end{eqnarray}
The prime indicates that in the corresponding curvature the term with $f_\mu$ has been omitted.

This finishes our discussion of the gauging of the $z=2$ Schr\"odinger algebra.

\section{The Null-reduction of  Conformal Gravity}

In this section we re-obtain the results on $z=2$ Schr\"odinger geometry  with arbitrary torsion obtained in the previous section by performing  a dimensional reduction of  conformal gravity  from $d+1$ to $d$ space-time dimensions along a null direction. We show that in this way one obtains the same transformation rules and the same expressions for the dependent spin-connections and special conformal gauge fields  as before.

Our starting point is conformal gravity in $d+1$ dimensions. Recall that the relativistic conformal algebra appends new generators to the translations and Lorentz transformations of the Poincar\'e algebra, namely dilatations and
special conformal transformations. When the algebra is gauged, the dilatations give rise to a gauge field $\hat b_M$ with associated
gauge parameter $\l_D$
 while the special conformal transformations are assigned a gauge field $\hat f_{M}{}^{A}$ and gauge parameters
$\l_K^{A}$.\,\footnote{Like in the Poincar\'e case we denote fields in $d+1$ dimensions with a hat.} Thus the full set of gauge fields is
\be
\{\hat{e}_M{}^A,\hat{\omega}_M{}{}^{AB},\hat b_M,\hat f_M{}^A\}\,.
\ee
It turns out that after imposing conventional constraints the spin-connection and special conformal gauge fields become dependent.
The transformation rules of the independent Vielbein and
dilatation gauge field are given by
\bea
\d\hat{e}_M{}^A&=&\l^{A}{}_{B}\hat e_{M}{}^{B}+\l_D\hat{e}_M{}^A~,
\\[.1truecm]
\d \hat b_M&=&\partial_M\l_D+\l_K^A\hat{e}_{MA}~.
\eea
Both gauge fields transform as covariant vectors under general coordinate transformations. Note that the dilatation gauge field transforms with a shift under the special conformal transformations and therefore can be gauged away by fixing the $K$-transformations. The expressions for the dependent spin-connections and special conformal gauge fields are given by
\bea
\label{confspinconn}
\hat{\omega}_{M}{}{}^{AB}(\hat{e},\hat b)&=&\hat{\omega}_{M}{}{}^{AB}(\hat{e})+2\hat{e}_{M}{}^{[A}\hat{e}^{B]N}\hat b_N~,\\[.2truecm]
\label{confspinconn2}
\hat{f}_{M}{}^A(\hat e,\hat b)&=&- \frac 1{d-1}\hat{{{\cal {R}}}}'_{M}{}^A+\frac 1 {2d(d-1)}\hat{e}_{M}{}^A\hat {{{\cal {R}}}}'~,
\eea
where $\hat {\cal {R}}_{MN}{}^{AB}$ is the Lorentz curvature of the conformal algebra and
\be
\hat {{{\cal {R}}}}'_{M}{}^A =\hat {\cal {R}}'_{MN}{}^{AB}\hat e^N{}_{B}~,\hskip 1.5truecm
\hat {\cal {R}}'= \hat e^M{}_{A}\hat {\cal {R}}'_{M}{}^A~.
\ee
The prime indicates that in the corresponding curvature the term with ${\hat f}_M{}^A$ has been omitted.

Using the same reduction Ansatz as in the NC case and splitting $\hat b_{M}=(b_{\mu},b_{v})$, we obtain the following transformation rules for the lower-dimensional fields:
\bea
\d \tau_{\m}&=&2\l_D\tau_{\m}~,
\\[.1truecm]
\d e_{\m}{}^{a}&=&\l^a{}_{b}\,e_{\m}{}^b + S^{-1}\l^a\tau_{\m}+\l_De_{\m}{}^a~,
\\[.1truecm]
\d m_{\m}&=&-\partial_\mu \zeta^v -S^{-1}\l_ae_{\m}{}^a~,
\\[.1truecm]
\label{dbm}
\d b_{\m}&=&\partial_{\m}\l_D+\l_K^ae_{\m a}+\l_K^{+}S^{-1}\tau_{\m}-\l_K^-Sm_{\m}~,
\\[.1truecm]
\label{dbu}
\d b_{\u}&=&\l_K^-S~,\\[.1truecm]
\d S&=& (\l+\l_D) S~.
\eea

From the last transformation rule it follows that gauge-fixing $S=1$ leads this time to a compensating Lorentz transformation with parameter
\be
\lambda_{\rm comp}= -\lambda_D\,.
 \ee
  The $S=1$ gauge-fixing is not sufficient to end up with the transformation rules of $z=2$ Schr\"odinger geometry as given in the previous section.
 The reason for this is that the null-reduction leads to as many $K$-transformations as components of ${\hat b}_M$ while in Schr\"odinger geometry we  have only a single $K$-transformation. This is related to the fact that the $z=2$ Schr\"odinger algebra
cannot be embedded into a higher-dimensional conformal algebra like the Bargmann algebra can be embedded into a higher-dimensional Poincar\'e algebra.
In order to obtain the same symmetries as $z=2$ Schr\"odinger geometry we need to impose a constraint that reduces the $d+1$ $K$-transformations to the single one corresponding to the Schr\"odinger algebra. To achieve this, we first gauge-fix $b_v=0$ which fixes $\lambda_K^-=0$. To gauge-fix another $d-1$ $K$-transformations we impose by hand the following constraint
\be\label{constraint}
{\cal R}_{0a}(H) = 0 \hskip .5truecm \rightarrow \hskip .5truecm b_a = - \tau_{0a}
\ee
This constraint has two effects. First of all, it fixes $d-1$ $K$-transformations, as can be seen from the following transformation rule:
\be
\d {\cal R}_{0a}(H)=2\lambda^b\tau_{ab}-\l^b{}_a{\cal R}_{0b}(H)-\lambda_D{\cal R}_{0a}(H)+2\l_{K\,a}.
\ee
Note that this gauge-fixing leads to the following compensation transformation:
\be\label{gflambdaK}
\lambda^{\rm comp}_{K\,a }= -\lambda^b\tau_{ab}\,.
\ee
At the same time, the gauge-fixing constraint \eqref{constraint} is a conventional constraint that allows us  to solve for the spatial components of the dilatation gauge field
as we did in the previous section. It is straightforward to check that after imposing the additional gauge-fixing condition  \eqref{constraint} and
identifying $\zeta^v = -\sigma\,, \lambda_K = \lambda_K^+$ we obtain precisely the transformation rules \eqref{transfSchr} of $z=2$ Schr\"odinger geometry as obtained in the previous section.

For completeness we also give the transformation rules of the projective inverses:
\bea
\d e^{\m}{}_a&=&-\l^b{}_ae^{\m}{}_b-\l_De^{\m}{}_a~,\\[.1truecm]
\d\tau^{\m}&=&-\l^ae^{\m}{}_a-2\l_D\tau^{\m}~.
\eea

We now consider the null-reduction of the dependent spin-connection and special conformal boost gauge fields.
The reduction of the spin-connection components is very similar to the NC  case.
We find that  the non-vanishing components  are given by
\bea
\hat{\omega}_{\mu}{}^{ab}(\hat e,\hat b)&\equiv& \omega_{\mu}{}^{ab}(e,\tau,m,b)=\mathring{\omega}_{\mu}{}^{ab}(e,\tau,m,b)-m_{\m}\tau^{ab}~,\\[.1truecm]
\hat{\omega}_{\mu}{}^{a+}(\hat e,\hat b)&\equiv &\omega_{\m}{}^a(e,\tau,m,b) = \mathring{\omega}_{\m}{}^a(e,\tau,m,b)~,\\[.1truecm]
\hat\omega_{\m}{}^{a-}(\hat e,\hat b)&=&e_{\m b}\tau^{ab}~,\\[.1truecm]
\hat\omega_{\m}{}^{-+}(\hat e,\hat b)&=&b_{\m}~,\\[.1truecm]
\hat\omega_{v}{}^{ab}(\hat e,\hat b)&=&\tau^{ab}~,
\eea
where $\mathring{\omega}_{\m}{}^{ab}(e,\tau,m,b)$ and $\mathring{\omega}_{\mu}{}^a(e,\tau,m,b)$ are the torsionless Schr\"odinger spin-connections, whose explicit
expressions are given in eqs.~\eqref{schr1} and \eqref{schr2}, respectively.

Next, we consider the null-reduction of
the gauge field of special conformal transformations $\hat{f}_{M}{}^A$, defined in eq.~\eqref{confspinconn2}.
After a straightforward calculation we find the following expressions:
\bea
\hat f_{\m}{}^a&=& -\frac 1{d-1} \left({\cal R}_{\mu b}(J^{ab})+m_{\m} {\cal D}_b\t^{ab}
		     - e_{\m b}{\cal D}_0\tau^{ab}
		    -\frac 1{2d}e_{\m}{}^a {\cal R}_{bc}(J^{bc})\right)
		 +\mathring{\omega}_{\mu}{}^b\tau_{ba}~, \label{hatfmua}
\\[.1truecm]
\hat f_{\mu}{}^+&=& \frac 1{d-1}\left({\cal R}'_{\mu a}(G^a)+{\cal R}'_{\mu 0}(D)\right) - \frac 1{2d(d-1)} m_{\m}{\cal R}_{ab}(J^{ab})~,\\[.1truecm]
\hat f_{\m}{}^-&=& \frac 1{2d(d-1)} \t_{\m}{\cal R}_{ab}(J^{ab})-\frac 1{d-1}e_{\mu b}{\cal D}_a\t^{ab}-\frac 1{d-1}m_{\m}\t^{ab}\t_{ab}~, \\[.1truecm]
\hat f_{\u}{}^a&=& \frac 1{d-1} {\cal D}_{b}\tau^{ab}~,\\[.1truecm]
\hat f_{\u}{}^+&=& \frac 1{2d(d-1)} {\cal R}_{ab}(J^{ab})~,\\[.1truecm]
\hat f_{\u}{}^-&=& \frac 1{d-1} \tau_{ab}\t^{ab}~\,,
\eea
where the gauge field $\hat f_{\m}{}^{+}$ is identified as  the single gauge field $f_{\m}$ of the reduced theory.
The covariant derivative ${\cal D}$ is defined exactly as in \eqref{eq:defcovdertorsion}, but this time with the spin-connections
$\mathring\omega_{\mu}{}^{ab}(e,\tau,m,b)$ and $\mathring\omega_{\mu}{}^{a}(e,\tau,m,b)$, see eqs.~\eqref{schr1} and \eqref{schr2}.
We observe that the component $\hat{f}_{\mu}{}^a$ contains a torsion term with an explicit appearance of the spin-connection
$\mathring{\omega}_{\mu}{}^a$. This is explained by the fact that $\hat{f}_{\mu}{}^a$ originally was a special conformal gauge field transforming as $\partial_{\mu}\lambda_{K}^a$ under the  special conformal transformations. However, due to the gauge-fixing
of those transformations, and in particular due to the compensating transformation given in eq.~\eqref{gflambdaK}, we obtain $\delta\hat{f}_{\mu}{}^a=\partial_{\mu}\lambda^b\tau_{ba}+\dots$,
which explains the last term in eq.~\eqref{hatfmua}.

\section{NC Gravity with Arbitrary Torsion}

In this section we will use  our results on Schr\"odinger geometry with arbitrary torsion, derived in the previous section, to construct the NC gravity equations of motion for arbitrary torsion by applying  the so-called conformal technique for the non-relativistic case \cite{Afshar:2015aku}. We will give complete results for $d=3$ only.

It turns out that only the NC equations of motion with zero torsion ($\tau_{0a}=0, \tau_{ab}=0$)  and with half-zero torsion ($\tau_{0a}=0, \tau_{ab}\ne 0$) are  invariant under central charge transformations.  However, the null-reduction by construction always leads to an answer that is invariant under central charge transformations. That is why we found that the on-shell null-reduction of the Einstein equations leads to NC gravity with zero torsion. The half-zero torsion condition, although consistent with invariance under central charge transformations,  has no clear causal structure and, as we saw above, does not follow from a null-reduction of General Relativity.

Applying the non-relativistic conformal technique \cite{Afshar:2015aku}, invariance under central charge transformations implies  that we only need to introduce a real compensating scalar $\varphi$ for dilatations and not a second one to compensate for the central charge transformations.  As was shown in \cite{Afshar:2015aku}, the SFT for this real scalar is given by\,\footnote{By a Schr\"odinger Field Theory (SFT) we mean a field theory that is invariant under the rigid Schr\"odinger symmetries, see, e.g.,\cite{Afshar:2015aku}.}
\begin{equation}\label{SFT1}
{\rm SFT1}\,:\hskip 1truecm \partial_0\partial_0\varphi=0\,,\hskip 1truecm \partial_a\varphi=0\,,\\[.1truecm]
\end{equation}
where the constraint $\partial_a\varphi$ is a consequence of the torsion condition $\tau_{0a}=0$. In the absence of this torsion condition, the equation $\partial_0\partial_0\varphi=0$ is not invariant under Galilean boosts. To make this equation invariant under Galilean boosts,
we introduce a second compensating scalar $\chi$ for central charge transformations. The important point is that under rigid Galilean boosts the spatial derivative of this compensating scalar $\chi$  transforms as \cite{Afshar:2015aku}
\begin{equation}
\delta \left(\partial_a\chi\right) = - {\footnotesize{\textrm{M}}}\lambda_a\,,
\end{equation}
where $\mass$ is a mass parameter.
Therefore, the lack of Galilean boost invariance of SFT1, see eq.~\eqref{SFT1}, in the absence of the constraint $\partial_a\varphi=0$ can be compensated by adding further terms to this equation containing $\partial_a\chi$. In this way one ends up with the following SFT \cite{Afshar:2015aku}:
\begin{equation}\label{SFT2}
{\rm SFT2}\,:\hskip 1truecm
\partial_0\partial_0\varphi -\frac{2}{{\footnotesize{\textrm M}}
} (\partial_0\partial_a\varphi )\partial_a\chi +
\frac{1}{{\footnotesize{\textrm M}}^2}
(\partial_a\partial_b\varphi)\partial_a\chi\partial_b\chi=0\,.
\end{equation}
The second compensating scalar breaks the invariance under central charge transformations.  The SFT2 theory corresponds to either the twistless-torsional case ($\tau_{0a}\ne0, \tau_{ab}= 0$) or the arbitrary torsion case ($\tau_{0a}\ne0, \tau_{ab}\ne 0$) case.

In a next step, we couple this SFT2 theory to the $z=2$ Schr\"odinger geometry with arbitrary torsion, we constructed in the previous section, by replacing all derivatives in \eqref{SFT2} by Schr\"odinger covariant ones. In order to do this, it proves convenient to use a definition of the dependent gauge field $f_\mu$ of special conformal transformations that differs from the one given in eq.~\eqref{fsolved}, by terms that transform covariantly under gauge transformations.
In order to avoid confusion, we will denote this dependent gauge field by $F_\mu$. It is defined as the solution of the following conventional constraints
\begin{align} \label{eq:defF}
 & \mathcal{R}_{0a}(G^a) - \frac{2}{\mass} (D^b \chi) \mathcal{R}_{0a}(J^a{}_b) + \frac{1}{\mass^2} (D^b\chi) (D^c \chi) \mathcal{R}_{ca}(J^a{}_b) +\frac{1}{\mass^3} (D^b\chi) (D_b\chi) (D^c\chi) D^a \tau_{ca} = 0 \,, \nonumber \\
 & \mathcal{R}_{0a}(D) = 0 \,,
\end{align}
where the curvatures are given by the expressions in eq.~\eqref{curvatureszis3}, with $f_\mu$ replaced by $F_\mu$. In particular one finds that $F_0 = \tau^\mu F_\mu$ is given by
\begin{align} \label{eq:defF0}
 F_0 &= \frac{1}{d-1} \Big(\mathcal{R}^\prime_{0a}(G^a) - \frac{2}{\mass} (D^b \chi) \mathcal{R}_{0a}(J^a{}_b) + \frac{1}{\mass^2} (D^b\chi) (D^c \chi) \mathcal{R}_{ca}(J^a{}_b) \,+ \nonumber \\ & \quad +\frac{1}{\mass^3} (D^b\chi) (D_b\chi) (D^c\chi) D^a \tau_{ca}\Big) \,.
\end{align}
With this definition, $F_\mu$ transforms as follows under the different gauge transformations:
\begin{align}
  \tau^\mu \delta F_\mu &= \tau^\mu\left(\partial_\mu \lambda_K + 2 \lambda_K b_\mu - 2 \lambda_D f_\mu + 2 \lambda^b \mathring{\omega}_\mu{}^c \tau_{bc} \right) - \frac{3}{\mass} \lambda^a (D^b\chi) D_0 \tau_{ab} \, + \nonumber \\ &\quad + \frac{1}{\mass^2} \lambda^a (D^b\chi) (D^c\chi) D_{c} \tau_{ab} \,, \nonumber \\
  e^\mu{}_a \delta F_\mu &= e^\mu{}_a \left(\partial_\mu \lambda_K + 2 \lambda_K b_\mu - 2 \lambda_D f_\mu + 2 \lambda^b \mathring{\omega}_\mu{}^c \tau_{bc} \right) - 2 \lambda^b D_0 \tau_{ba} \,.
\end{align}
The first step in coupling the equations of the SFT2 theory \eqref{SFT2} to the $z=2$ Schr\"odinger geometry with arbitrary torsion consists of replacing all derivatives in the left-hand-side of \eqref{SFT2} by Schr\"odinger covariant ones. This leads to the expression
\begin{equation}\label{total}
D_0D_0\varphi -\frac{2}{{\footnotesize{\textrm M}}
} (D_0D_a\varphi )D_a\chi +
\frac{1}{{\footnotesize{\textrm M}}^2}
(D_aD_b\varphi)D_a\chi D_b\chi \,,
\end{equation}
where the covariant derivatives are given by\,\footnote{For the special case $\tau_{ab}=0$  the expressions were already given in \cite{Afshar:2015aku}.}
\begin{eqnarray}
D_0D_0\varphi &=& \tau^\mu\big(\partial_\mu D_0\varphi +b_\mu D_0\varphi + \mathring{\omega}_\mu{}^aD_a\varphi +  F_\mu\varphi\big)\,,\\[.1truecm]
D_0D_a\varphi &=& \tau^{\mu}\big(\partial_{\mu}D_a\varphi - \mathring{\omega}_{\mu a}{}^{b}D_b\varphi + \mathring{\omega}_\mu{}^b \tau_{ba} \varphi\big)\,,\\[.1truecm]
D_aD_b\varphi &=& e_{a}^{\mu}\big(\partial_{\mu}{D_{b}{\varphi}}  - \mathring{\omega}_{\mu b}{}^{c}D_c{\varphi} + \mathring{\omega}_\mu{}^c \tau_{cb} \varphi \big) \,,\\[.1truecm]
D_0\varphi &=& \tau^\mu\big (\partial_\mu -b_\mu\big)\varphi\,,\hskip 1.5truecm D_a\varphi = e_a{}^\mu\big (\partial_\mu -b_\mu\big)\varphi\,,\\[.1truecm]
D_a\chi &=&  e_{a}^{\mu}\big(\partial_{\mu}{\chi} - \mass m_{\mu}\big)\,.
\end{eqnarray}
Note that the second covariant time derivative $D_0D_0\varphi$ of $\varphi$ contains the time-component $\tau^\mu F_\mu$ of the dependent special conformal gauge field gauge field $F_\mu$ given as a solution of eqs.~\eqref{eq:defF}.

The expression \eqref{total} can not be used yet as the starting point for defining a Schr\"odinger covariant equation, as it is not yet invariant under
local boost transformations. Indeed, one finds that its variation under boosts is given by
\begin{align}\label{variation}
&  - \frac{2}{\mass^2} (D^a \chi) (D^b \chi) \lambda_a \tau_b{}^c D_c \varphi - \frac{1}{\mass^2} (D^a \chi) (D_a \chi) \lambda^b \tau_b{}^c D_c \varphi \,.
\end{align}
We expect that this variation can be cancelled by adding further terms to the expression \eqref{total} via an iterative procedure but we did not yet find a closed answer in arbitrary dimensions. However, for the special case of $d=3$, the calculation simplifies significantly and the variation \eqref{variation}
can be cancelled by adding two extra terms to the expression \eqref{total}. As a result, we find that the following equation is Schr\"odinger invariant in $d=3$:
\begin{align} \label{eq:total2}
 & D_0 D_0 \varphi - \frac{2}{\mass} (D_0 D_a \varphi) (D^a \chi) + \frac{1}{\mass^2} (D_a D_b \varphi) (D^a \chi) (D^b \chi) \, - \nonumber \\ & \qquad  - \frac{1}{\mass^3} (D^a \chi) (D_a \chi) (D^b \chi) \tau_b{}^c D_c \varphi + \frac{1}{4 \mass^4} (D^a \chi) (D_a \chi) (D^b \chi) (D^c \chi) \tau_{b}{}^d \tau_{cd} \varphi = 0 \,.
\end{align}
To present the field equations, it is convenient to introduce the following boost invariant connection for spatial rotations
\begin{align}
  \Omega_\mu{}^{ab} &= \mathring{\omega}_\mu{}^{ab} + H_\mu{}^{ab} \,,
\end{align}
with the covariant tensor $H_\mu{}^{ab}$ given by
\begin{align}
  H_\mu{}^{ab} = \frac{1}{\mass} D_\mu \chi \tau^{ab} + \frac{2}{\mass} \left(e_\mu{}^c D^{[a} \chi + D^c \chi e_\mu{}^{[a} \right) \tau_c{}^{b]} - \frac{2}{\mass^2} \tau_\mu D_c \chi D^{[a} \chi \tau^{b]c} \,.
\end{align}
With this definition, the curvature tensor
\begin{align}
  \mathcal{R}_{\mu\nu}{}^{ab}(\Omega) = 2 \partial_{[\mu} \Omega_{\nu]}{}^{ab} - 2 \Omega_{[\mu}{}^{ac} \Omega_{\nu] c}{}^b
\end{align}
is boost invariant and related to $\mathcal{R}_{\mu\nu}{}^{ab}(J)$ via
\begin{align}
  \mathcal{R}_{\mu\nu}{}^{ab}(\Omega) = \mathcal{R}_{\mu\nu}{}^{ab}(J) + 2 D_{[\mu} H_{\nu]}{}^{ab} - 2 H_{[\mu}{}^{ac} H_{\nu]c}{}^b \,.
\end{align}
Since $\mathcal{R}_{\mu\nu}{}^{ab}(\Omega)$ is boost invariant, one can consistently impose
\begin{align} \label{Reoms1}
  \mathcal{R}_{0b}{}^{ba}(\Omega) = 0 \,, \qquad \mathcal{R}_{ac}{}^{cb}(\Omega) = 0
\end{align}
as two of the NC field equations. Under boost transformations, the first equation in \eqref{Reoms1} transforms to the second one while the second one is invariant. These two equations are the extension to arbitrary torsion of the last two zero torsion NC equations given in eq.~\eqref{eomzerotorsion}. The extension to arbitrary torsion of the first zero torsion equation $R_{0a}(G^a)=0$ given in \eqref{eomzerotorsion} can  be found by imposing in eq.~\eqref{eq:total2} the gauge-fixing conditions
\begin{align}\label{gaugefix}
\varphi=1\,,\hskip 2truecm \chi=0\,,
\end{align}
fixing the dilatations and central charge transformations, respectively. After substituting the expressions of the dependent Schr\"odinger gauge fields we derived in the previous two sections and using the other two torsional equations of motion \eqref{Reoms1},
we find that this third torsional NC equation is given by
\begin{eqnarray}\label{third}
&&\tfrac{1}{2} {\cal R}^\prime_{0a}(G^a) - \tau^\mu\mathring{\omega}_\mu{}^a b_a -2 ({\cal D}_0b_a)m^a - ({\cal D}_a b_b)m^a m^b -(b_c-\sfrac 14 m^c\tau_{cd})\tau^b{}_{c}m^am_am^b - \nn\\[.2truecm]
	&& -\sfrac 12 m^am_am^bD_{c}\tau_b{}^c-m^be_a{}^{\nu}(\tau^{\mu}+\sfrac 12 m^ce_c{}^{\mu})(2 D_{[\mu} H_{\nu]}{}^{ab} - 2 H_{[\mu}{}^{ac} H_{\nu]c}{}^b)=0\,,
\end{eqnarray}
where
\begin{equation}
{\cal D}_{\mu}b_a=\partial_{\mu}b_a-\mathring{\omega}_{\mu a}{}^{b}b_b+\mathring{\omega}_{\mu}{}^b\tau_{ab}\,,
\end{equation}
$b_0$ is gauge-fixed to zero,  and, after the gauge-fixing \eqref{gaugefix}, $H_\mu{}^{ab} $ is given by
\begin{equation}
H_\mu{}^{ab} = -m_{\mu} \tau^{ab} - {2} \left(e_\mu{}^c m^{[a}  + m^c  e_\mu{}^{[a} \right) \tau_c{}^{b]} - {2} \tau_\mu m_c  m^{[a}  \tau^{b]c}\,.
\end{equation}
Note that $H_{\mu}{}^{ab}$ vanishes identically for the special case that the torsion $\tau_{ab}$ is zero.

This finishes our discussion of NC gravity with arbitrary torsion in three dimensions whose equations can be found in eqs.~\eqref{Reoms1} and \eqref{third}.

\section{Conclusions}
\label{conclusions}

In this paper we applied two complementary techniques, gauging and null-reduction, to construct Newton-Cartan geometry and its conformal extension, $z=2$ Schr\"odinger geometry, with arbitrary torsion.The gauging technique has the advantage that it makes the symmetries
resulting from the construction manifest. The null-reduction technique has the advantage that the construction is algorithmic and can easily be generalized to other cases as well. We explained why the null-reduction technique does not yield NC gravity with arbitrary torsion and showed, in three space-time dimensions, how equations of motion with arbitrary torsion can be obtained by applying the non-relativistic conformal method \cite{Afshar:2015aku} using a SFT with two real compensating scalars: one compensating scalar $\varphi$ for the dilatations and one compensating scalar $\chi$ for the central charge transformations. This compensating technique leads to one of the equations of motion of torsional NC gravity, see eq.~\eqref{third}. This singlet equation is the one that
contains the Poisson equation of the Newton potential. The other two equations, see eq.~\eqref{Reoms1},  followed
by formulating them in terms of the curvature of a boost-invariant connection.

It would be interesting to extend the results of this paper to the supersymmetric case and apply the null-reduction technique to supergravity theories. The case of $d=3$ should lead to a generalization  of the  off-shell 3D NC supergravity  constructed in \cite{Bergshoeff:2015uaa,Bergshoeff:2015ija} to the case of arbitrary torsion. More interestingly, one can also take $d=4$ and construct 4D NC supergravity thereby obtaining, after gauge-fixing, the very first supersymmetric generalization of 4D Newtonian gravity. An intriguing feature of the 3D case is that the
Newtonian supergravity theory contains both a Newton potential as well as a dual  Newton potential. In analogy to the 3D case, we expect that in the supersymmetic case the Newton potential will not occur in the same representation as introduced by Newton. It would be interesting to see which representations of the Newton potential would occur in the 4D case and investigate whether this could have any physical effect.

\paragraph{Acknowledgements.}
E.A.B. and J.R. gratefully acknowledge support from the Simons Center for Geometry and Physics, Stony Brook University at which part of the research for this paper was performed during the workshop {\sl Applied Newton-Cartan Geometry}. E.A.B. and J.R. also thank the Galileo Galilei Institute in Firenze for the stimulating atmosphere during the workshop {\sl Supergravity, what next?}, where this work was initiated. E.A.B.  wishes to thank the University of Vienna for its hospitality. The
work of A.Ch. was  supported by the H2020 Twinning project No. 692194, "RBI-T-WINNING".


\begin{thebibliography}{99}





\bibitem{Christensen:2013lma}
  M.~H.~Christensen, J.~Hartong, N.~A.~Obers and B.~Rollier,
  ``Torsional Newton-Cartan Geometry and Lifshitz Holography,''  Phys.\ Rev.\ D {\bf 89} (2014) 061901  doi:10.1103/PhysRevD.89.061901  [arXiv:1311.4794 [hep-th]].  


\bibitem{Geracie:2014nka}
  M.~Geracie, D.~T.~Son, C.~Wu and S.~F.~Wu,
  ``Spacetime Symmetries of the Quantum Hall Effect,''  Phys.\ Rev.\ D {\bf 91} (2015) 045030  doi:10.1103/PhysRevD.91.045030  [arXiv:1407.1252 [cond-mat.mes-hall]].  




\bibitem{Luttinger:1964zz}
  J.~M.~Luttinger,
  ``Theory of Thermal Transport Coefficients,''  Phys.\ Rev.\  {\bf 135} (1964) A1505.  doi:10.1103/PhysRev.135.A1505  






\bibitem{Gromov:2014vla}
  A.~Gromov and A.~G.~Abanov,
  ``Thermal Hall Effect and Geometry with Torsion,''  Phys.\ Rev.\ Lett.\  {\bf 114} (2015) 016802  doi:10.1103/PhysRevLett.114.016802  [arXiv:1407.2908 [cond-mat.str-el]].  


\bibitem{Geracie:2014mta}
  M.~Geracie, S.~Golkar and M.~M.~Roberts,
  ``Hall viscosity, spin density, and torsion,''  arXiv:1410.2574 [hep-th].  


\bibitem{Geracie:2016dpu}
M.~Geracie, K.~Prabhu and M.~M.~Roberts,
``Physical stress, mass, and energy for non-relativistic matter,''
JHEP {\bf 1706} (2017) 089
doi:10.1007/JHEP06(2017)089
[arXiv:1609.06729 [hep-th]].


\bibitem{Duval:1984cj}
  C.~Duval, G.~Burdet, H.~P.~Kunzle and M.~Perrin,
  ``Bargmann Structures and Newton-cartan Theory,''  Phys.\ Rev.\ D {\bf 31} (1985) 1841.  doi:10.1103/PhysRevD.31.1841  



\bibitem{Julia:1994bs}
  B.~Julia and H.~Nicolai,
  ``Null Killing vector dimensional reduction and Galilean geometrodynamics,''
  Nucl.\ Phys.\ B {\bf 439} (1995) 291
  [hep-th/9412002].

\bibitem{Jensen:2014aia}
  K.~Jensen,
  ``On the coupling of Galilean-invariant field theories to curved spacetime,''  arXiv:1408.6855 [hep-th].  

\bibitem{Bekaert:2014bwa}
  X.~Bekaert and K.~Morand,
  ``Connections and dynamical trajectories in generalised Newton-Cartan gravity I. An intrinsic view,''  J.\ Math.\ Phys.\  {\bf 57} (2016) no.2,  022507  doi:10.1063/1.4937445  [arXiv:1412.8212 [hep-th]].  




\bibitem{Festuccia:2016awg}
  G.~Festuccia, D.~Hansen, J.~Hartong and N.~A.~Obers,
  ``Torsional Newton-Cartan Geometry from the Noether Procedure,''  Phys.\ Rev.\ D {\bf 94} (2016) no.10,  105023  doi:10.1103/PhysRevD.94.105023  [arXiv:1607.01926 [hep-th]].


\bibitem{Andringa:2010it}
  R.~Andringa, E.~Bergshoeff, S.~Panda and M.~de Roo,
  ``Newtonian Gravity and the Bargmann Algebra,''  Class.\ Quant.\ Grav.\  {\bf 28} (2011) 105011  doi:10.1088/0264-9381/28/10/105011  [arXiv:1011.1145 [hep-th]].  


\bibitem{Afshar:2015aku}
  H.~R.~Afshar, E.~A.~Bergshoeff, A.~Mehra, P.~Parekh and B.~Rollier,
  ``A Schr\"odinger approach to Newton-Cartan and Ho\v rava-Lifshitz gravities,''  JHEP {\bf 1604} (2016) 145  doi:10.1007/JHEP04(2016)145  [arXiv:1512.06277 [hep-th]].  


\bibitem{Bergshoeff:2014uea}
  E.~A.~Bergshoeff, J.~Hartong and J.~Rosseel,
  ``Torsional Newton-Cartan geometry and the Schr\"odinger algebra,''
  Class.\ Quant.\ Grav.\  {\bf 32} (2015) no.13,  135017
  [arXiv:1409.5555 [hep-th]].


\bibitem{Hartong:2015zia}
  J.~Hartong and N.~A.~Obers,
  ``Ho\v rava-Lifshitz gravity from dynamical Newton-Cartan geometry,''  JHEP {\bf 1507} (2015) 155  doi:10.1007/JHEP07(2015)155  [arXiv:1504.07461 [hep-th]].  


\bibitem{Bergshoeff:2015uaa}
  E.~Bergshoeff, J.~Rosseel and T.~Zojer,
  ``Newton-Cartan (super)gravity as a non-relativistic limit,''  Class.\ Quant.\ Grav.\  {\bf 32} (2015) no.20,  205003  doi:10.1088/0264-9381/32/20/205003  [arXiv:1505.02095 [hep-th]].  

\bibitem{Bergshoeff:2015ija}
  E.~Bergshoeff, J.~Rosseel and T.~Zojer,
  ``Newton-Cartan supergravity with torsion and Schr\"odinger supergravity,''  JHEP {\bf 1511} (2015) 180  doi:10.1007/JHEP11(2015)180  [arXiv:1509.04527 [hep-th]].  






\end{thebibliography}
\end{document}